# Computational Study of Rarefied Gas Flow and Heat Transfer in Lid-driven Cylindrical Cavities

Mengbo Zhu (朱孟波)[1], Ehsan Roohi[1,*], Amin Ebrahimi[2,*]

1. State Key Laboratory for Strength and Vibration of Mechanical Structures, International Center for Applied Mechanics (ICAM), School of Aerospace Engineering, Xi'an Jiaotong University (XJTU), Xianning West Road, Beilin District, Xi'an 710049, China

2. Department of Materials Science and Engineering, Faculty of Mechanical, Maritime and Materials Engineering, Delft University of Technology, Mekelweg 2, 2628 CD, Delft, The Netherlands

* Corresponding author: E.Roohi@xjtu.edu.cn (E. Roohi), A.Ebrahimi@tudelft.nl (A. Ebrahimi)

## Abstract

The gas flow characteristics in lid-driven cavities are influenced by several factors, such as cavity geometry, gas properties, and boundary conditions. In this study, the physics of heat and gas flow in cylindrical lid-driven cavities with various cross-sections, including fully or partially rounded edges, is investigated through numerical simulations using the direct simulation Monte Carlo (DSMC) and the discrete unified gas kinetic scheme (DUGKS) methods. The thermal and fluid flow fields are systematically studied for both constant and oscillatory lid velocities, for various degrees of gas rarefaction ranging from the slip to the free-molecular regimes. The impact of expansion cooling and viscous dissipation on the thermal and flow fields, as well as the occurrence of counter-gradient heat transfer (also known as anti-Fourier heat transfer) under non-equilibrium conditions, are explained based on the results obtained from numerical simulations. Furthermore, the influence of the incomplete tangential accommodation coefficient on the thermal and fluid flow fields is discussed. A comparison is made between the thermal and fluid flow fields predicted in cylindrical cavities and those in square-shaped cavities. The present work contributes to the advancement of micro/nano-electromechanical systems (MEMS/NEMS) by providing valuable insights into rarefied gas flow and heat transfer in lid-driven cavities.

## 1. Introduction

The physics of rarefied gas flow in cavities is a critical area of research within the field of fluid dynamics, as it is relevant to a wide range of engineering and scientific applications, including micro- and nano-scale devices, vacuum systems, fuel cells, atomic force microscopes, space propulsion, and atmospheric entry of space vehicles [1–5]. The mean free path of the gas molecules in such systems is often comparable with or exceeds the characteristic length scale of the fluid flow involved [6], leading to notable distinctions between gas and liquid flow in these systems [7–10]. As a result, the flow deviates from thermodynamic equilibrium, leading to non-equilibrium effects such as velocity slip, temperature jump, and non-Maxwellian velocity distributions of the gas molecules [6].

The Knudsen number (Kn), which is a dimensionless parameter, is defined as the ratio of the mean free path of the gas molecules, $\lambda$, to a characteristic length scale of the flow, $L$. The greater the Knudsen number, the more significant the deviation from thermodynamic equilibrium [11]. The Knudsen number is commonly employed to classify gas flows into continuum (Kn < $10^{-3}$), slip ($10^{-3}$ < Kn < $10^{-1}$), transition ($10^{-1}$ < Kn < 10), and free-molecular (Kn > 10) regimes [12]. However, it should be noted that the classification of different Knudsen number regimes is inherently empirical, and the limits of these regimes may differ for fluid flows in complex geometries [13–15]. In continuum flow, the frequency of molecular collisions is sufficiently high, and the gas behaves as a continuum; thus, the Navier-Stokes-Fourier equations can describe the flow. As the frequency of molecular collisions decreases with increasing the Knudsen number, the kinetic behavior of the individual molecules must be taken into account to describe the flow. Under such non-equilibrium conditions, conventional continuum models based on the Navier-Stokes-Fourier equations are no longer valid, and kinetic models must be utilized to solve the Boltzmann equation to describe the gas flow [16].

The direct simulation Monte Carlo (DSMC) method [17] is a widely utilized computational technique for simulating gas flows. As a kinetic method, the DSMC method employs stochastic processes to simulate the motion of individual gas molecules. The DSMC method is based on the numerical solution of the Boltzmann equation, which describes the evolution of the distribution function of the gas molecules. However, its computational expense increases when simulating gas flows close to the continuum regime, as it requires smaller time-step and grid sizes than the molecular collision time and mean free path [18]. Recently, the discrete unified gas-kinetic scheme (DUGKS) [19–22] has been developed to solve multiscale problems using the gas kinetic equation. The efficacy and efficiency of the DUGKS method have been verified, particularly in the near-continuum regime [23–26].

The characteristics of gas flow in lid-driven cavities are affected by several factors, including the geometry of the cavity, the properties of the gas, and the boundary conditions. A considerable amount of research has been conducted on the subject of rarefied gas flow and heat transfer in lid-driven cavities with geometries that include sharp corners, such as triangular [27], rectangular [28–33], and trapezoidal [34]. These studies have analyzed the impact of several factors on the thermal and flow fields, such as expansion cooling, viscous dissipation, incomplete surface accommodation, and pressure variations. Additionally, these studies have also reported the occurrence of unique physical phenomena, like counter-gradient heat transfer (also known as anti-Fourier heat transfer), in lid-driven cavities with geometries that include sharp corners under non-equilibrium conditions. Despite the extensive research conducted on the subject, and numerous studies on the Couette flow problem in cylindrical coordinates [35–39], relatively less research has been conducted on heat and gas flow in lid-driven cylindrical cavities. Furthermore, previous studies have mostly focused on steady-state gas flow behavior in cavities driven by constant lid velocity. Therefore, further research is necessary to improve the current understanding of gas behavior in lid-driven cylindrical cavities to design and optimize functional devices and systems that are both efficient and sustainable.

The present work focuses on describing the physics of heat and gas flow in cylindrical lid-driven cavities of various cross-sections, including those with fully or partially rounded edges, utilizing the direct simulation Monte Carlo (DSMC) and the discrete unified gas kinetic scheme (DUGKS) methods. The effects of both constant and oscillatory lid velocities on the thermal and fluid flow fields are described. The manuscript is organized as follows: Section 2 presents the cylindrical cavity geometries and boundary conditions, Section 3 describes the numerical techniques used, Section 4 presents the results of the thermal and gas flow fields, and finally, the conclusion is provided in Section 5.

## 2. Problem description

The present work investigates the thermal and fluid flow behavior of argon, a monoatomic gas, in two different cavities, as depicted in Figure 1. The first cavity, designated as P-Cavity, is a cylindrical cavity with a flat top lid, with the horizontal lid located at $y = R \cdot \sin(\pi/4)$, where $R$ represents the radius of the cavity, as illustrated in Figure 1(a). The second cavity, referred to as C-Cavity (see Figure 1(b)), is a cylindrical cavity with a 90° arc lid. Both cavities have a diameter of 1 μm and are centered at the coordinates $O(x, y) = (0, 0$ μm$)$. The solid boundaries of the cavity are maintained at a constant temperature of $T_w = 300$ K. The gas flow within the cavities is induced by the harmonic movement of the lid. Specifically, the lid oscillates at a constant frequency $\psi$, and the magnitude of its velocity is defined as follows:

$$U_{\text{lid}} = U_{\text{M}} \cdot \cos(\psi t), \tag{1}$$

where, $U_{\text{M}}$ is the maximum amplitude of the oscillating lid velocity magnitude, and $t$ is the time.

Thermal and gas flow behavior in cavities with constant lid velocity are also considered in the present work as a special case, where the oscillation frequency $\psi$ is equal to zero. The oscillatory gas flow behavior in cavities can be characterized by the cavity geometry, the Knudsen number and the Strouhal number. The Knudsen number (Kn) and the Strouhal number (St) are defined as follows [40]:

$$\text{Kn} = \frac{\lambda}{D} = \frac{\mu}{n_0 D}\left(\frac{\pi}{2m\text{K}_\text{B}T_\text{w}}\right)^{-1/2}, \tag{2}$$

$$\text{St} = \frac{\psi D}{v_\text{m}}, \tag{3}$$

where, $\lambda$ is the molecular mean free path, $D$ is the cavity diameter, $\text{K}_\text{B}$ is the Boltzmann constant, $T_\text{w}$ is the wall temperature, $m$ is the molecular mass, and $v_\text{m}$ is the most probable molecular velocity that is determined as follows [40]:

$$v_\text{m} = \left(\frac{2\text{K}_\text{B}T_\text{w}}{m}\right)^{-1/2}. \tag{4}$$

The computational grid configuration used in the simulations is also shown in Figure 1 for different cavity geometries.

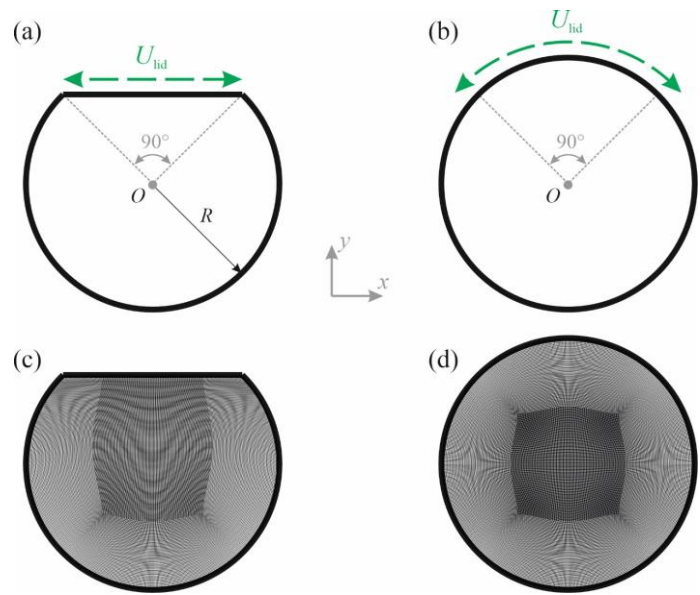

Figure 1. Geometries of the cavities studied in the present work. (a) P-Cavity, cylindrical cavity with a flat top lid, (b) C-Cavity, cylindrical cavity with a 90° arc lid, and (c) and (d) the computational grid configurations used in the simulations. Nonuniform quadrilateral grids were used to discretize the computational domain.

## 3. Methods

### 3.1. Direct simulation Monte Carlo (DSMC)

The DSMC method is a useful computational approach for simulating rarefied gas flows as it can accurately capture the behavior of individual gas molecules, even at low densities. The present work utilizes the direct simulation Monte Carlo (DSMC) method, a probabilistic, particle-based approach for simulating the kinetic behavior of gas particles. This method is used to approximate the solution of the Boltzmann kinetic equation that is defined as follows [41–43]:

$$\frac{\partial}{\partial t}(f) + \boldsymbol{\xi} \cdot \frac{\partial}{\partial r}(f) = \int_{-\infty}^{\infty} \int_{0}^{4\pi} (f^* f_1^* - f f_1) c_r s_c \, d\Omega dc_1, \tag{5}$$

where, $f$ and $f_1$ are the velocity distribution functions, $\boldsymbol{\xi}$ is the molecular velocity vector, $\boldsymbol{r}$ is the position vector, $t$ is the time, $n$ is the number density, $\boldsymbol{f}^*, \boldsymbol{f}_1^*$ are the corresponding velocity distribution after collision, $\boldsymbol{c}_r$ is the relative velocity vector of pre- and post-collision, $s_c$ is the collision cross-section, and $\Omega$ is the unit solid angle. In the DSMC method, the following splitting scheme is applied to the distribution function at $t_k$ to obtain the solution at $t_{k+1}$ [18];

$$\boldsymbol{f}[t_k + \delta t, \boldsymbol{x}(t_k), \boldsymbol{\xi}(t_k + \delta t)] = S_Q^{\delta t, h}\{\boldsymbol{f}[t_k, \boldsymbol{x}(t_k), \boldsymbol{\xi}(t_k)]\}, \tag{6}$$

$$\boldsymbol{f}[t_k + \delta t, \boldsymbol{x}(t_k + \delta t), \boldsymbol{\xi}(t_k + \delta t)] = S_D^{\delta t}\{\boldsymbol{f}[t_k + \delta t, \boldsymbol{x}(t_k), \boldsymbol{\xi}(t_k + \delta t)]\}, \tag{7}$$

where, $\delta t$ is the time step and operators $S_Q^{\delta t, h}$ and $S_D^{\delta t}$ are the DSMC numerical algorithms approximating the collision and free molecular motion terms in the Boltzmann equation respectively.

In this context, the Maxwell-type diffuse-specular condition is applied to model gas-solid surface interactions as collisions between gas molecules and surface material atoms [44]. The tangential accommodation coefficient ($\alpha_t$) was employed to represent the degree of specular-diffuse reflection of the surface, with values ranging from 0 to 1. A fully diffuse surface ($\alpha_t = 1$) is assumed to result in reflected molecules that are in a Maxwell-Boltzmann equilibrium distribution. Conversely, a fully specular boundary ($\alpha_t = 0$) is equivalent to a binary collision of hard spheres, where the particles retain their velocity in the tangent direction but change direction in the normal direction. The probability density function and distribution function are given by [45]:

$$S(\boldsymbol{v}_r|\boldsymbol{n}_w) = \alpha \frac{|\boldsymbol{v}_r \cdot \boldsymbol{n}_w|}{2\pi(\mathcal{R}T_r)^2} exp\left(-\frac{v_r^2}{2\mathcal{R}T_r}\right) + (1-\alpha)\,\delta_3\big(\boldsymbol{v}_r - (\boldsymbol{v}_i - 2(\boldsymbol{v}_i \cdot \boldsymbol{n}_w)\boldsymbol{n}_w)\big), \tag{8}$$

$$\boldsymbol{f}(\boldsymbol{v}_r) = \alpha \frac{n_r}{(2\pi\mathcal{R}T_r)^{3/2}} exp\left(-\frac{v_r^2}{2\mathcal{R}T_r}\right) + (1-\alpha)\,\boldsymbol{f}\big(\boldsymbol{r}_w, \boldsymbol{v}_r - (\boldsymbol{v}_i - 2(\boldsymbol{v}_i \cdot \boldsymbol{n}_w)\boldsymbol{n}_w, t\big), \tag{9}$$

where, $\boldsymbol{v}$ is the velocity vector, $\boldsymbol{n}_w$ is the unit vector normal to the surface, $\alpha$ is the accommodation coefficient, $\mathcal{R}$ is the gas constant, $\delta_3$ is the generalized Dirac delta function, $n_r$ is the number density, and $T_r$ is the surface temperature. The subscripts i and r indicate the incident and reflected parameters respectively. The effects of tangential accommodation coefficient $\alpha_t$ on the thermal and fluid flow fields in lid-driven cylindrical cavities are studied in the present work for the cases with Kn = 1.

The DSMC method works by dividing the flow domain into a number of small control volumes called simulation cells. Inside each cell, the motion of individual gas molecules is simulated using a Monte Carlo algorithm. The algorithm generates random positions and velocities for the gas molecules based on their distribution function at the start of the simulation. As the simulation progresses, the algorithm updates the positions and velocities of the gas molecules based on their interactions with the boundaries of the simulation cell. These interactions are modelled using collision models which incorporate the properties of the gas and the boundary conditions of the flow.

In the present study, the direct simulation Monte Carlo (DSMC) simulations were conducted utilizing the open-source solver, dsmcFoam+ [46]. The dsmcFoam+ solver has been widely employed to model rarefied gas flow and heat transfer at micro and nano scales [15, 47–51]. The working fluid selected for this investigation is argon, with a molecular mass of $m = 6.63{\times}10^{-26}$ kg and a molecular diameter of $d = 4.17 \times 10^{-10}$ m. Argon molecules possess three translational degrees of freedom, but do not possess any rotational degrees of freedom. The viscosity-temperature index was set to $\omega = 0.81$, with a reference temperature of $T_{ref} = 273$ K. The variable-hard-sphere (VHS) intermolecular collision model was employed in the simulations, and collision pairs were chosen according to the standard no-time-counter (NTC) method.

In the simulations, a maximum grid size of $0.1\lambda$ was used and the number of particles per cell (PPC) was set to 30. The time step size, $\Delta t$, was chosen to be $0.01\Delta t_c$, which is significantly smaller than the mean collision time $\Delta t_c$. The transient adaptive sub-cells (TAS) method was implemented to optimize computational efficiency, which adjusts sub-cell sizes based on the number of particles per sub-cell (PPSC). To ensure steady-state conditions were achieved, all simulations were run for a minimum of $10^8$ time-steps.

## 3.2. Discrete unified gas kinetic scheme (DUGKS)

The discrete unified gas kinetic scheme (DUGKS) is a deterministic method that is efficient and accurate for simulating rarefied gas flows with a wide range of Knudsen numbers, including those with complex geometries, while DSMC is a stochastic method that is particularly well-suited for simulating rarefied gas flows with high Knudsen numbers but can become computationally expensive with large number of simulated molecules. Thus, the choice between DUGKS and DSMC depends on the specific requirements of the problem at hand. DUGKS shares the advantages of the unified gas kinetic scheme (UGKS) and the lattice Boltzmann method [52].

DUGKS is a numerical method for solving the kinetic equation of rarefied gas flows. The method is based on the unified gas kinetic scheme (UGKS) and the Bhatnagar–Gross–Krook (BGK)–Shakhov model, which includes both the Navier-Stokes-Fourier and the Burnett equations as special cases. The DUGKS method discretizes the UGKS model using a finite-volume method and a two-step time-marching scheme. The resulting algorithm is highly efficient and accurate and is able to handle a wide range of rarefied gas flow problems, including those with complex geometry and boundary conditions [23]. Additionally, the DUGKS method has been shown to be more robust and less sensitive to numerical errors than other methods for solving the kinetic equation of rarefied gas flows [53]. Details of the DUGKS method are reported comprehensively elsewhere [19, 20, 23, 53], and thus are not repeated here.

An open-source DUGKS solver, dugksFoam [53], was employed to construct DUGKS simulations in the present work. A grid size of 0.2$\lambda$ with 28 Gauss-Hermite discrete velocity points was employed in the simulations, and the time-step size $\Delta t$ was adapted automatically based on the Courant-Friedrich-Lewy number (CFL = $\xi \cdot \Delta t / \Delta x$) to achieve a CFL less than 0.5, where $\xi$ is the molecular velocity magnitude, and $\Delta x$ is the computational grid size.

## 4. Model verification

A comprehensive study was performed to assess the impact of the computational grid size, time-step size ($\Delta t$), and number of particles per cell (PPC) on the numerical predictions of direct simulation Monte Carlo (DSMC) simulations, and the results are presented in Figure 2. For this study, the gas flow in the P-Cavity with a constant lid velocity and a Knudsen number of 0.5 was considered. The results in Figure 2 are displayed along a horizontal line at $y/D = 0.35$ for temperature and a vertical line at $x/D = 0$ for normalized pressure and Mach number (Ma = $U/\gamma \mathcal{R} T$, where $U$ is the gas velocity magnitude, $\gamma$ is the ratio of the specific heats, $\mathcal{R}$ is the gas constant, and $T$ is the temperature). As seen in Figure 2(a–c), among the various grid sizes tested (0.5$\lambda$, 0.1$\lambda$, and 0.02$\lambda$), a grid size of 0.1$\lambda$ was determined to be sufficient to accurately capture the thermal and gas flow characteristics.

The time-step size in DSMC simulations should be small enough to decouple the molecular movement from molecular collision. To determine the optimal time-step size ($\Delta t$), the mean collision time ($\Delta t_c$) was calculated as follows [54]:

$$\Delta t_c = \lambda \left( \frac{2 K_B T_w}{m} \right)^{-1/2},$$                     (10)

where, $\lambda$ is the molecular mean free path, $K_B$ is the Boltzmann constant, $T$ is the temperature, and $m$ is the molecular mass. The time-step size in the simulations was set to $0.01\Delta t_c$, based on the results presented in Figure 2(d–f). The influence of PPC on the numerical predictions is shown in Figure 2(g–i), leading to the selection of a PPC value of 30 for the DSMC simulations. Figure 3 shows the effect of computational grid size on the numerical predictions obtained from the DUGKS simulations. The results showed that a grid size of $0.2\lambda$ is adequate for accurately simulating the thermal and gas flow characteristics using the DUGKS model.

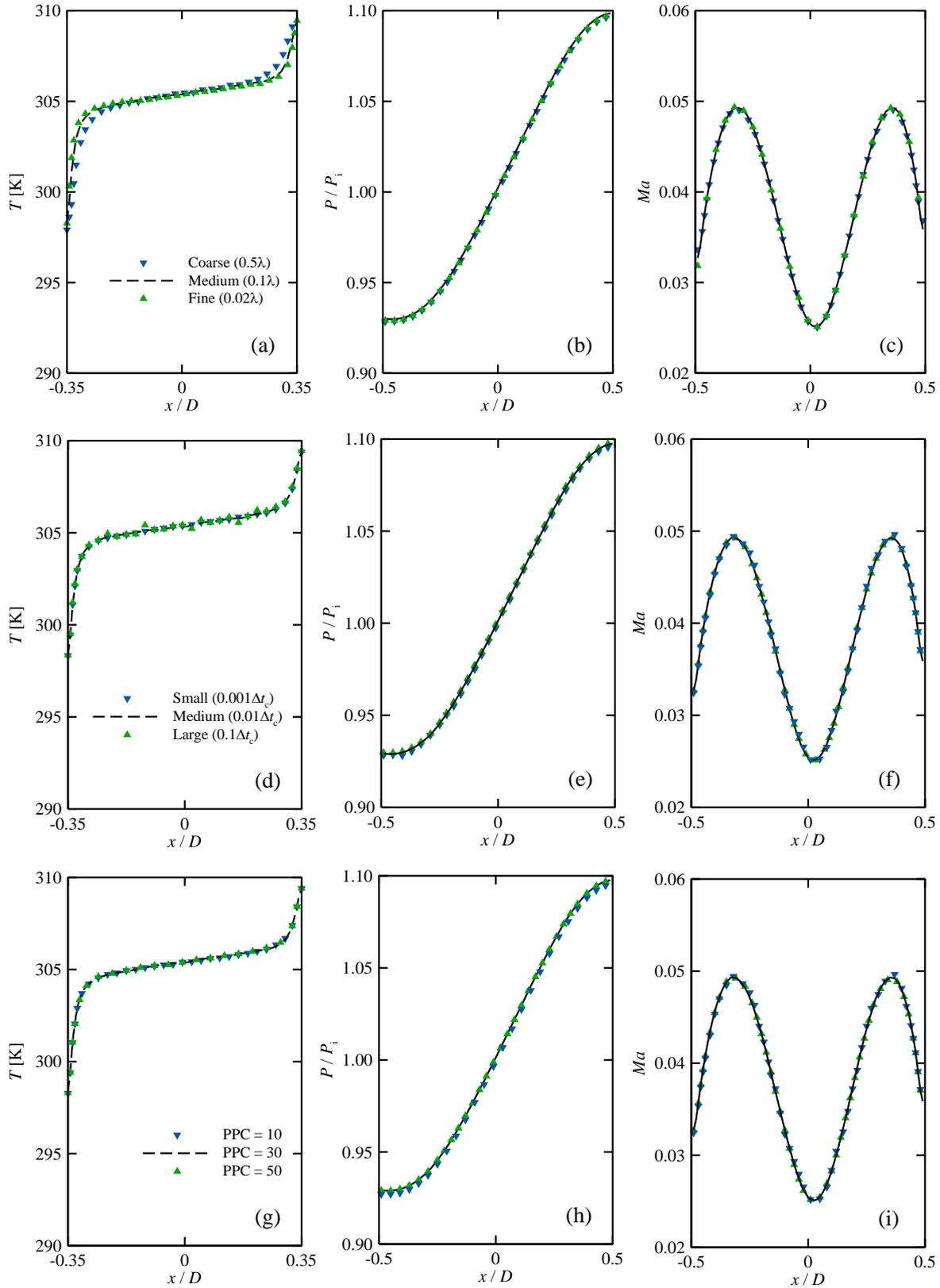

Figure 2. The effects of the computational gird size, the time-step size ($\Delta t$), and the number of particles per cell (PPC) on the predicted thermal and gas flow fields obtained from the DSMC simulations. Data was obtained from a simulation of the P-Cavity with a constant lid velocity and Kn = 0.5. The pressure values are normalized with respect to the initial absolute pressure in the cavity ($P_i = n K_B T_w$).

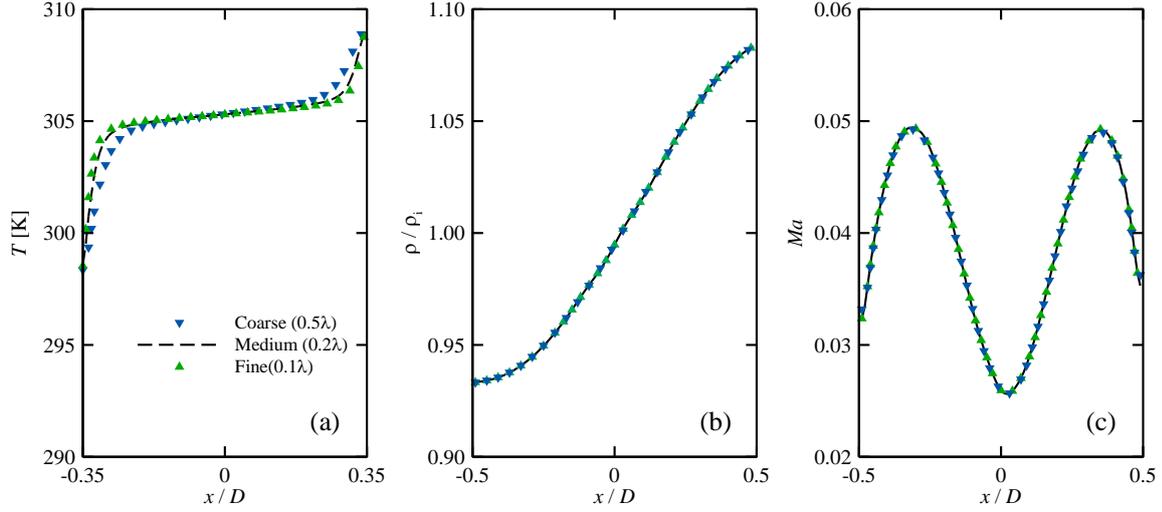

Figure 3. The influence of the computational grid size on the numerical predictions obtained from the DUGKS simulations. Data was obtained from a simulation of the P-Cavity with a constant lid velocity and Kn = 0.5. The density values are normalized with respect to the initial gas density in the cavity ($\rho_i = m\lambda^{-1}(2\pi d^2)^{-1/2}$).

A comparison of the numerically predicted thermal and fluid flow fields using the DUGKS and DSMC models is presented in Figure 4 for the case of P-Cavity with a constant lid velocity and a Knudsen number of $10^{-1}$. The results of both models are in close agreement with each other, with a deviation of less than 0.1%. Specifically, the values predicted for the normalized velocity components ($u/U_{\text{lid}}$ at $x/D = 0.5$ and $v/U_{\text{lid}}$ at $y/D = 0.5$) and temperature ($T$ at $y/D = 0.5$) show a high degree of consistency between the two models. This suggests that the DUGKS and DSMC models are both capable of accurately predicting the thermal and rarefied gas flow fields in lid-driven cylindrical cavities.

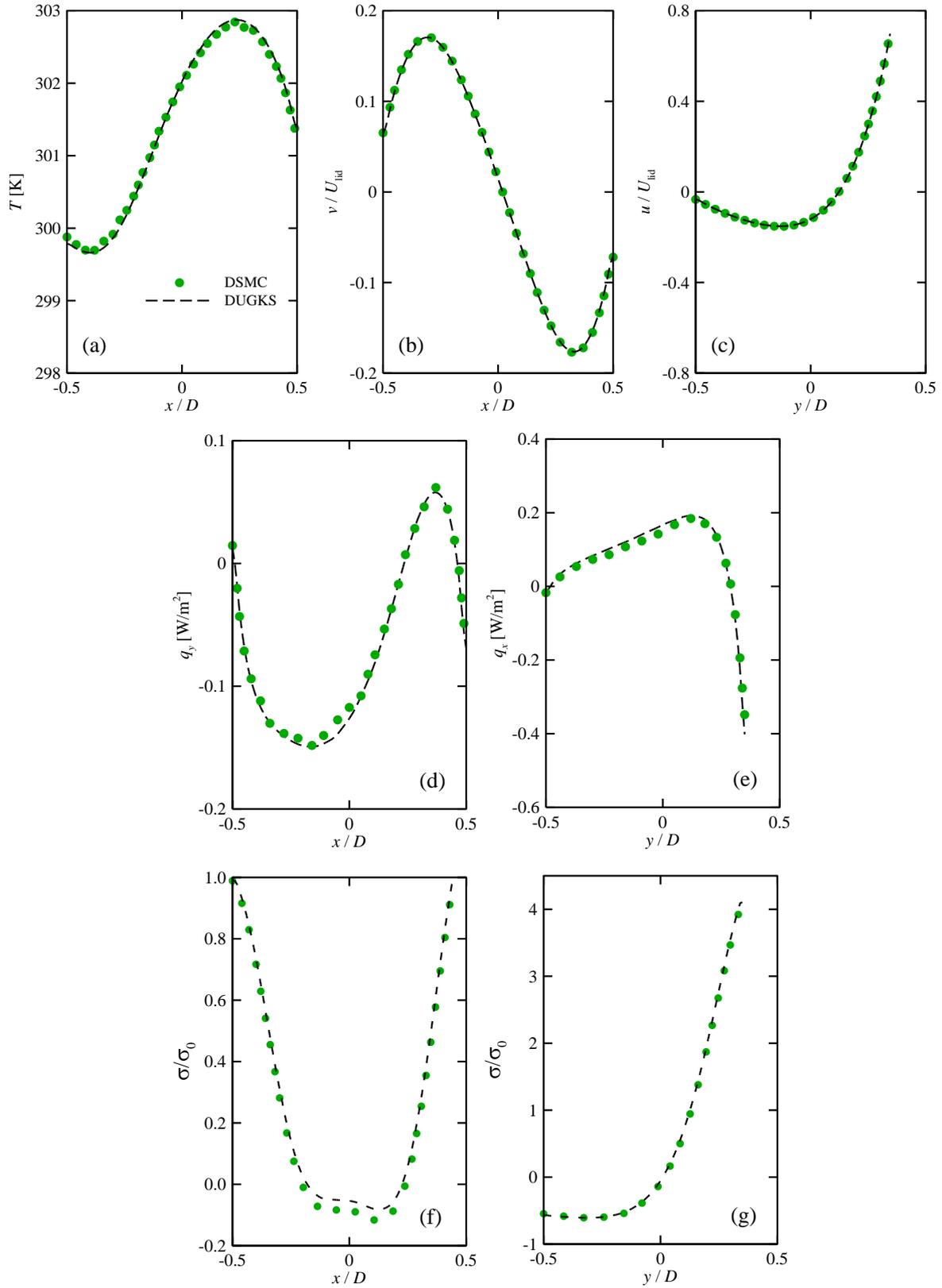

Figure 4. Comparison of the results obtained from the DUGKS simulations with those obtained from the DSMC simulations for gas flow in the P-Cavity with a constant lid velocity and Kn = 0.1.

The selection of the model was based on its performance in specific conditions, with the DUGKS model being employed for low Knudsen number values of $10^{-2}$ and $10^{-1}$ due to its computational efficiency and accuracy in modelling oscillatory flows, as demonstrated in the literature [55–57]. The DSMC model was employed for higher Knudsen number values of $10^{-1}$, 1, and 10 due to its superior performance in accurately modelling gas-surface interactions.

## 5. Results and discussion

In this section, the numerical predictions of gas flow in cylindrical cavities are presented. Thermal and gas flow characteristics in cavities with different geometries and constant lid velocity (*i.e.*, St = 0) are predicted using both the DSMC and DUGKS models for a wide range of Knudsen numbers ($10^{-2} \leq$ Kn $\leq 10$) encompassing the slip, transition, and free-molecular flow regimes. Additionally, thermal and gas flow characteristics in cavities with oscillatory lid movement are predicted using the DUGKS model in the late slip-early transition rarefaction regime (Kn = $10^{-1}$) for different Strouhal numbers ranging between 1 and 10.

The flow system considered in the present work can be regarded as a "time-period steady-state" [56, 57] in which the flow-related variables oscillate periodically with a constant frequency and the flow field reaches a steady-state condition after several oscillation periods. In other words, the flow variables are time-dependent, but the pattern of the flow variables repeats over time, and the flow variables at any given point in space reach a steady state after a certain number of periods. The maximum velocity within the cavity is observed at the cavity lid at the instant when the dimensionless time $t/t_s$ equals zero, where $t_s$ is the period of the oscillation and $\frac{t}{t_s} = \psi t / \pi$ [40]. Conversely, the maximum velocity in the opposite direction occurs at $t/t_s = 0.5$. The results obtained exhibit symmetry in one complete period $t_s$; hence, only the results from the first half of the period are presented here.

## 5.1. Characteristics of the flow field

### 5.1.1. Constant lid velocity

Figure 5 shows the predicted velocity field in cylindrical cavities with different shapes and Knudsen numbers, driven by a constant lid velocity. The movement of the cavity lid at a constant velocity drives the gas flow in the cavity, resulting in the formation of a vortex. The results indicate that the average gas velocity in the P-Cavity is generally lower than that in the C-Cavity. Moreover, the average gas velocity decreases by about 40% in the P-Cavity and about 16% in the C-Cavity with increasing the Knudsen number and reaches a minimum at a Knudsen number of approximately 1. However, the average gas velocity in the cavity remains almost constant for Knudsen numbers greater than 1. As the Knudsen number increases, the mean free path of molecules in the gas becomes larger, resulting in a decrease in the frequency of molecular collisions. A decrease in the frequency of molecular collisions leads to an increase in the average production of peculiar velocities, $\overline{c_x \cdot c_y}$. In other words, the less frequent the molecular collisions are, the less quickly the peculiar velocities will be damped out. Hence, as the production of peculiar velocities increases, the velocity of the gas molecules becomes more disordered, leading to more significant fluctuations in the velocity field.

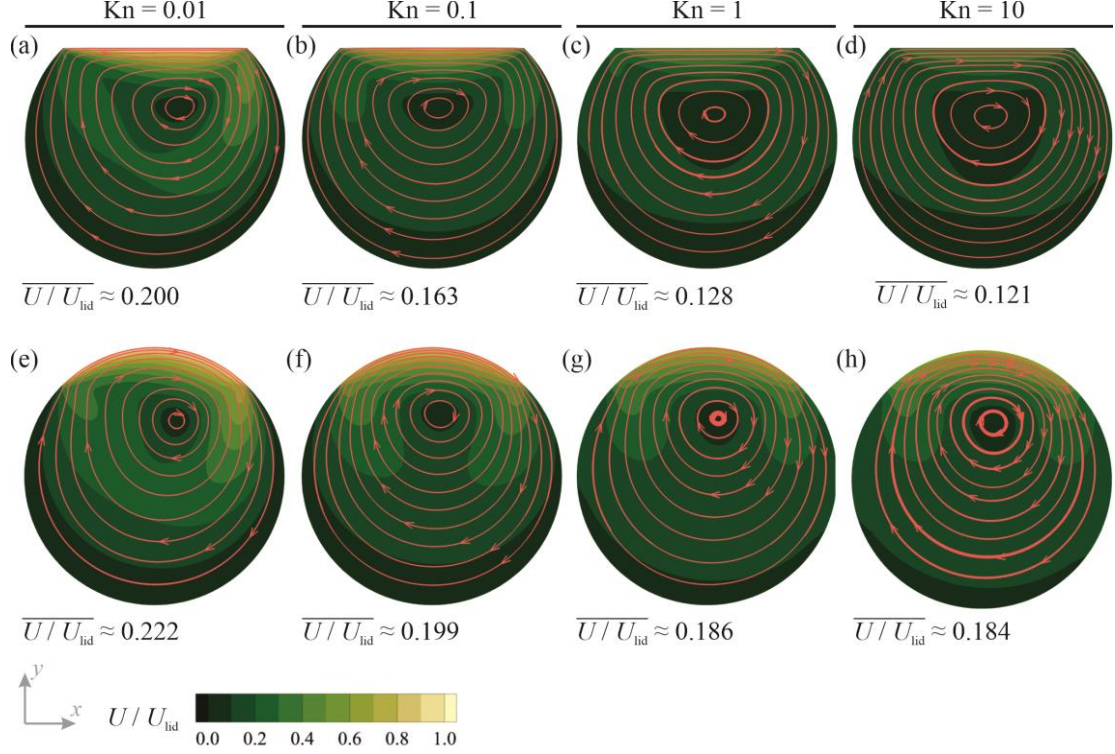

Figure 5. Contours of the normalized velocity magnitude ($U/U_{\text{lid}}$) overlayed by streamlines for different cavity shapes and Knudsen numbers. The velocity magnitude of the cavity lid $U_{\text{lid}}$ is constant (*i.e.*, St = 0). Data corresponding to Knudsen numbers of 0.01 and 0.1 were acquired using the DUGKS model, while data for Knudsen numbers of 1 and 10 were obtained from the DSMC model.

The dynamic viscosity of a gas is generally related to the mean free path of the molecules and their collision frequency. As the mean free path increases with increasing the Knudsen number, the collision frequency decreases, which leads to a decrease in the dynamic viscosity. However, the precise relationship between the Knudsen number and the dynamic viscosity can be complex and depends on factors such as gas composition and temperature [58]. At very low Knudsen numbers (*i.e.* in the continuum regime), the dynamic viscosity is independent of the Knudsen number and depends only on the gas properties and temperature [17]. As the Knudsen number increases and the gas becomes more rarefied, the dynamic viscosity decreases due to the reduced frequency of molecular collisions. Accordingly, an increase in the Knudsen number results in a decrease in the rate of momentum transfer through the fluid and a subsequent decrease in the shear stress.

Figure 6 shows the predicted location of the vortex center as a function of Knudsen number for cavities with different geometries. It appears that the center of the vortex in the P-Cavity moves away from the moving boundary and shifts slightly towards the cavity center with increasing the Knudsen number. Similar behavior has been reported for rarefied gas flow in lid-driven square cavities [5, 28, 29, 56]. As the Knudsen number increases from $10^{-2}$ to $10^{-1}$, the vortex center in the C-Cavity moves towards the moving boundary and shifts towards the cavity center. However, the vortex center moves away from the moving boundary with further increase in the Knudsen number. The results indicate that changes in the horizontal position of the vortex center become insignificant for Knudsen numbers beyond 1 in cavities regardless of their geometry. The variation in the position of the vortex center as a function of the Knudsen number is ascribed to the rarefaction phenomenon caused by a reduction in intermolecular collisions. The corresponding reduction in momentum transfer among molecules results in a decrease in viscosity. This is in line with previous studies, which indicate that at extremely small Knudsen numbers, the movement of the vortex center is downwards when the Reynolds number is raised [59].

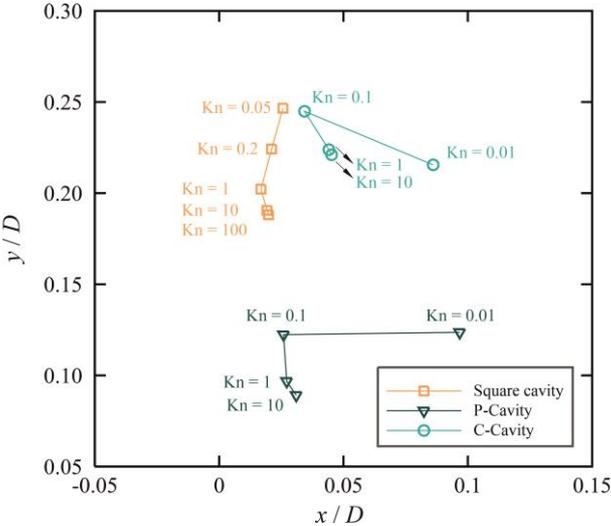

Figure 6. The locus of the vortex center in P-Cavity (triangles), C-Cavity (circles) and square cavity (squares) as a function of the Knudsen number. The velocity magnitude of the cavity lid $U_{lid}$ is constant (*i.e.*, St = 0). Data for the square cavity is taken from RezapourJaghargh *et al*. [5].

Figure 7 shows the normalized velocity profiles for different cavity shapes and Knudsen numbers. The results indicate that an increase in the Knudsen number results in an increase in the velocity slip at the solid boundaries. However, for both cavity shapes, the slip velocity remains nearly constant when the Knudsen number increases from 1 to 10. When the Knudsen number exceeds $10^{-1}$, the velocity field exhibits symmetry about the $y$-axis crossing the cavity center. In lid-driven cavities, the velocity difference between the moving lid and the adjacent gas generates a velocity gradient near the lid, resulting in shear stress. The magnitude of shear stress is directly proportional to the velocity difference between the moving lid and the adjacent gas. Therefore, the magnitude of shear stress is greatest in regions close to the lid and decreases moving away from the lid. However, the magnitude of shear stress is significantly reduced as the Knudsen number increases, which is consistent with the results reported by John *et al.* [29] for rarefied gas flow in lid-driven square cavities. This can be attributed to the reduced frequency of molecular collisions and the decrease in the rate of momentum transfer through the gas.

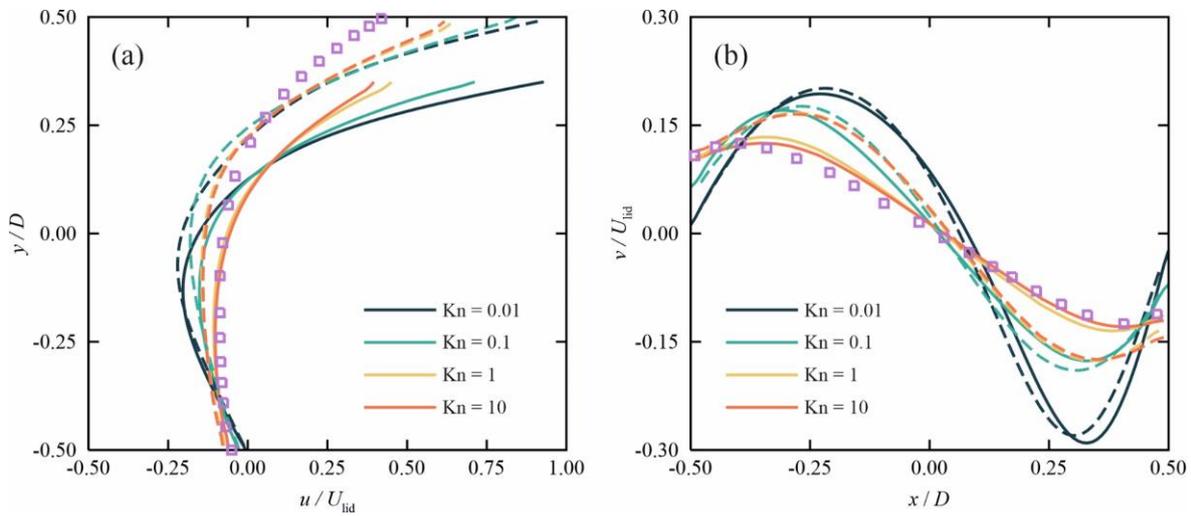

Figure 7. Normalized velocity profiles for different cavity shapes (solid lines: P-Cavity, dashed lines: C-Cavity, and symbols: square cavity) and Knudsen numbers. (a) the profile of horizontal velocity component $u$ along a vertical line, and (b) the profile of vertical velocity component $v$ along a horizontal line crossing the cavity center. The velocity magnitude of the cavity lid $U_{\text{lid}}$ is constant (*i.e.*, St = 0). Data for the square cavity is taken from John *et al.* [29] for Kn = 1.

The behavior of gas molecules in proximity to solid surfaces has a significant impact on the macroscopic behavior of the flow [60, 61]. When the Knudsen number is low (Kn ≪ 1), molecular collisions occur frequently, and the gas behaves as a continuum. As a result, viscous forces dominate the molecular interactions with solid surfaces. However, when the Knudsen number increases, molecular collisions occur less frequently, and the molecular interactions with solid surfaces are dominated by the ballistic collisions of gas molecules with the surface. The behavior of gas molecules near solid surfaces is significantly influenced by the tangential accommodation coefficient $\alpha_t$, which is a dimensionless quantity that ranges between 0 (fully specular reflection) and 1 (fully diffusive reflection) [38, 62–64]. The value of the tangential accommodation coefficient is generally dependent on the type of gas and the conditions of the wall surface [64–67]. The tangential accommodation coefficient defines the ratio of the momentum accommodation coefficient for tangential momentum to that for normal momentum and plays a critical role in describing the ability of gas molecules to transfer momentum with a solid surface and determining the slip velocity in rarefied flows.

Figure 8 shows the normalized velocity profiles for different cavity shapes and tangential accommodation coefficients at Kn = 1. When the tangential accommodation coefficient $\alpha_t$ is close to unity, which corresponds to diffusive reflection, the slip velocity at the moving lid is small. A decrease in the tangential accommodation coefficient results in an increase in the slip velocity at the moving lid, which in turn lead to a decrease in the shear stress near the lid. This decrease in shear stress can significantly affect the flow behavior, such as the development of vortices and the mixing of fluids. Figure 9 shows the change in the locus of the vortex center in cylindrical cavities as a function of the tangential accommodation coefficient. A decrease in the tangential accommodation coefficient results in an increase in the slip velocity at the moving lid. This increase in slip velocity leads to a decrease in the shear stress near the lid, affects the pressure field in the cavity and results in a less efficient momentum transfer [68]. Hence, the vortex center moves downwards and towards the geometric center of the cavity with decreasing the tangential accommodation coefficient. The results suggest that changes in the tangential accommodation coefficient lead to larger variations in the vortex center's location in cylindrical cavities than in square cavities.

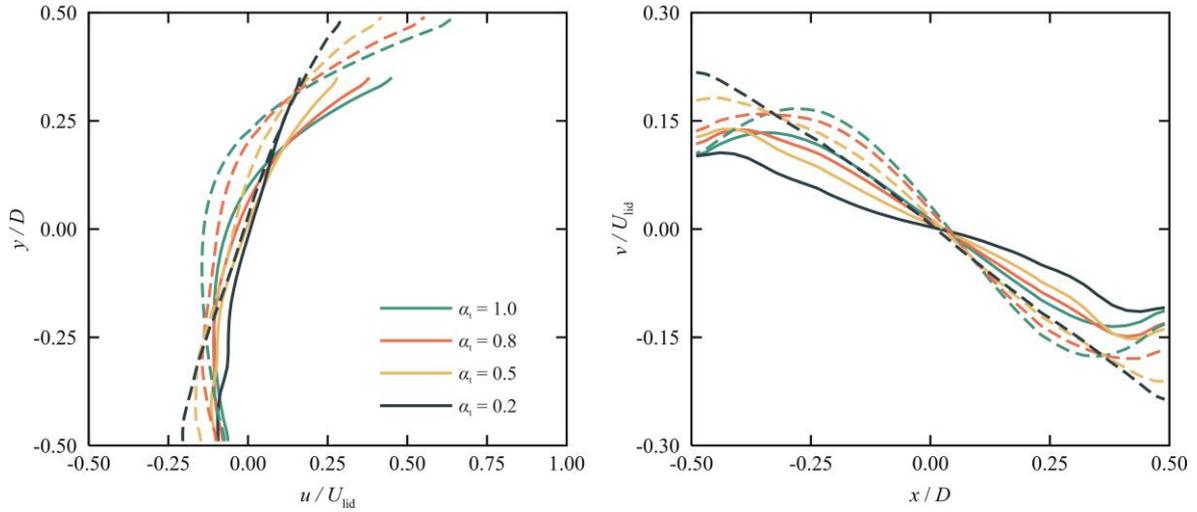

Figure 8. Normalized velocity profiles for different cavity shapes (solid lines: P-Cavity, dashed lines: C-Cavity) and tangential accommodation coefficients. (a) the profile of horizontal velocity component $u$ along a vertical line, and (b) the profile of vertical velocity component $v$ along a horizontal line crossing the cavity center. The velocity magnitude of the cavity lid $U_{lid}$ is constant (*i.e.*, St = 0) and the Knudsen number is equal to 1.

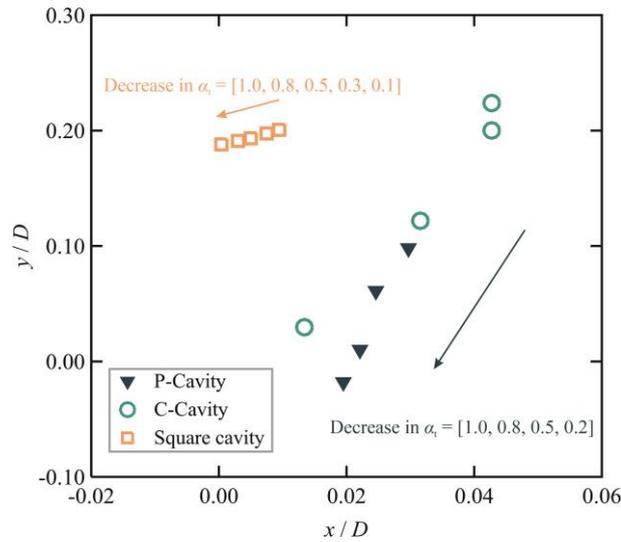

Figure 9. The locus of the vortex center in P-Cavity (triangles), C-Cavity (circles) and square cavity (squares) as a function of the tangential accommodation coefficient $\alpha_t$. The velocity magnitude of the cavity lid $U_{lid}$ is constant (*i.e.*, St = 0) and the Knudsen number is equal to 1. Data for the square cavity is taken from John *et al*. [30] for Kn = 1.

### 5.1.2. Oscillatory lid velocity

Figure 10 shows the velocity field in cylindrical cavities with different shapes and Strouhal numbers at $t/t_s = 0$, driven by an oscillatory lid velocity. For these cases, the Knudsen number is equal to $10^{-1}$. The results show that an increase in the lid oscillation frequency results in a decrease in the average fluid velocity in the cavity. When the lid oscillation frequency is low, *i.e.*, St = 1 and 2, the streamlines in both cavity shapes are closed throughout an oscillation quarter-period, indicating a vortex flow structure in the cavity. However, an increase in the lid oscillation frequency from St = 2 to St = 5 and 10 results in open streamlines. This higher frequency of lid oscillation destroys the original vortex flow, and a source-sink flow replaces it at the upper region near the lid. Moreover, the flow in the cavity no longer follows the cavity shape. Furthermore, increasing the lid oscillation frequency results in an increase in the number of sources and sinks in the cavity. Similar behavior has been observed in oscillatory rarefied gas flow in square and rectangular cavities [40, 56, 69].

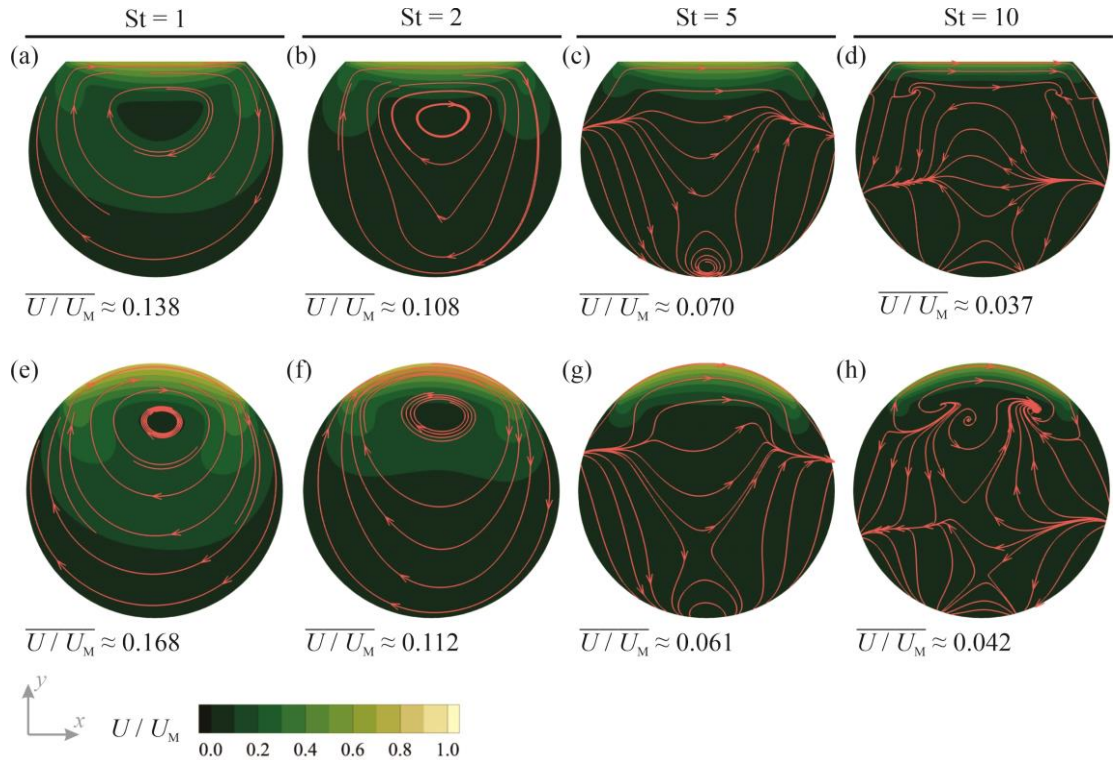

Figure 10. Contours of the normalized velocity magnitude ($U/U_M$) overlayed by streamlines for different cavity shapes and Strouhal numbers. The results are shown at $t/t_s = 0$. The Knudsen number is equal to $10^{-1}$.

Figure 11 shows the time evolution of the velocity field in the P-Cavity for various lid oscillation frequencies. At St = 1, the gas flow in the cavity is in sync with the lid oscillation, and the streamlines are closed. However, as the frequency of the lid oscillation increases, the phase lag relative to the lid also increases. This causes the gas flow far away from the lid in the cavity to become out of phase with the lid, resulting in open streamlines. Figure 12 shows the evolution of the velocity field in the C-Cavity for different Strouhal numbers, and it exhibits a similar behavior to the P-Cavity.

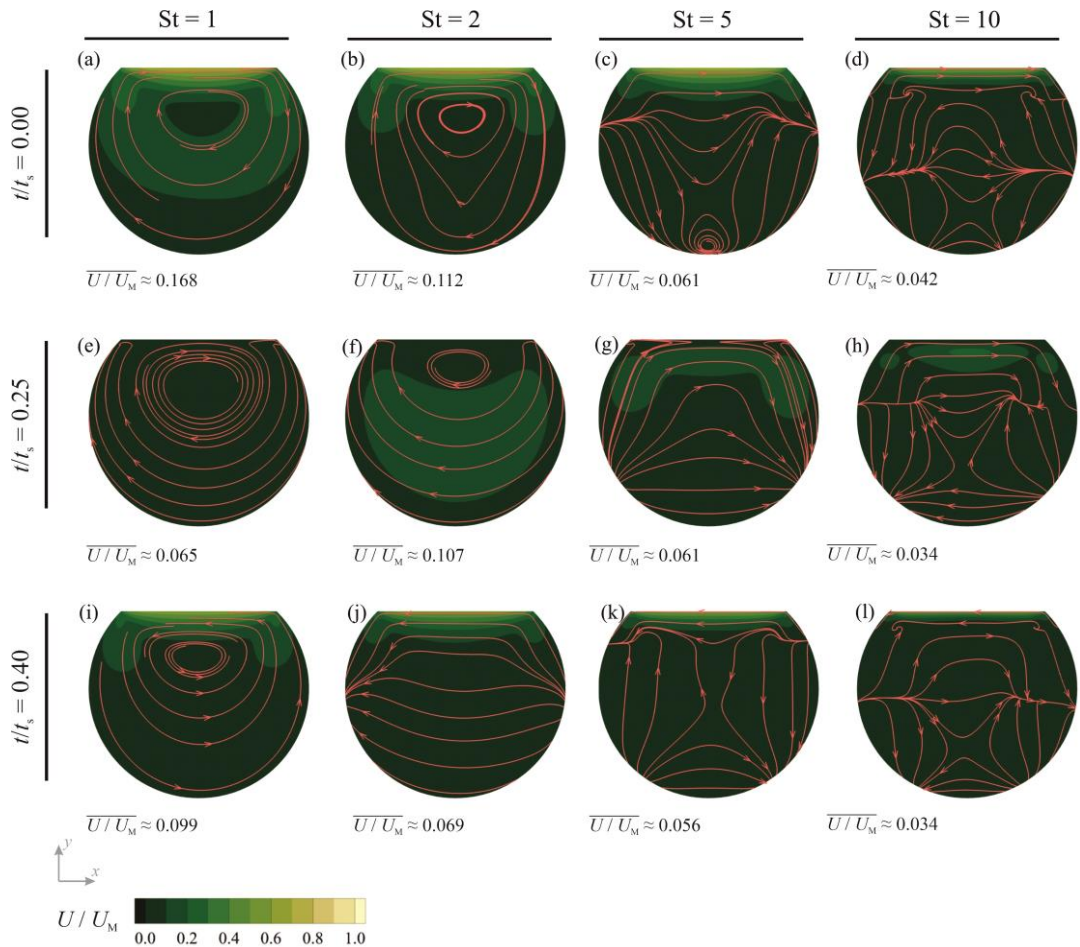

Figure 11. Contours of the normalized velocity magnitude ($U/U_M$) overlayed by streamlines in the P-Cavity at different time instances for different Strouhal numbers. The lid velocity magnitude reaches its minimum at $t/t_s = 0.25$. The Knudsen number is equal to $10^{-1}$.

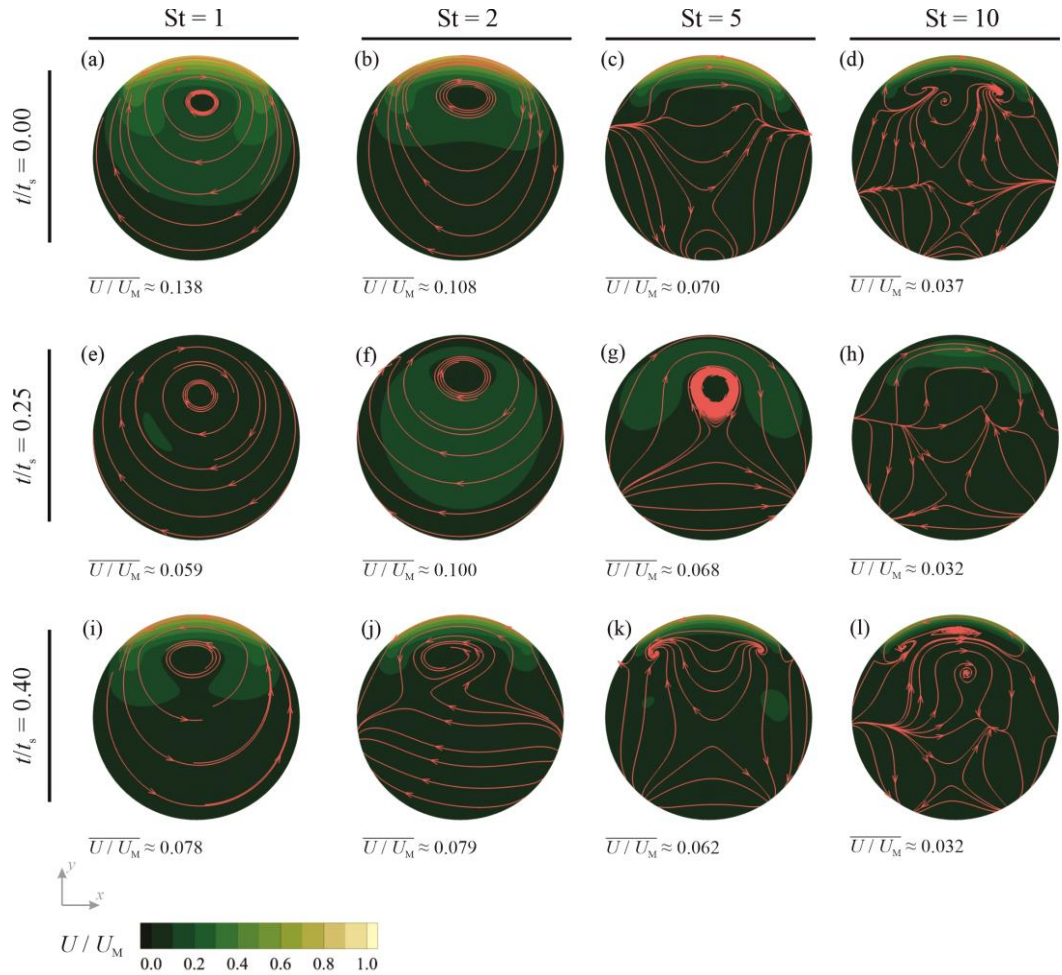

Figure 12. Contours of the normalized velocity magnitude ($U/U_M$) overlayed by streamlines in the C-Cavity at different time instances for different Strouhal numbers. The lid velocity magnitude reaches its minimum at $t/t_s$ = 0.25. The Knudsen number is equal to $10^{-1}$.

The profiles of normalized velocity in cylindrical cavities for different Strouhal numbers are shown in Figure 13. As the lid oscillation frequency increases, the slip velocity at the lid also increases, resulting in a larger magnitude of shear stress in regions close to the lid. The slip velocity at the moving lid decreases by 26.8% in the P-Cavity and by 39.6% in the C-Cavity with an increase in the Strouhal number from 0 to 10. However, the slip velocities at the lid of a square cavity are generally greater than those observed in cylindrical cavities, as shown in Figure 13. The results from Figure 13(a) reveal that the gas horizontal velocity decreases as moving away from the lid. This decrease in velocity is more pronounced at higher Strouhal numbers, leading to an increased velocity gradient and hence an increase in shear stress near the lid at higher lid oscillation frequencies. For the vertical velocity, its magnitude decreases as the Strouhal number increases until it reverses direction at St = 5, corresponding to the downward bending of streamlines. Moreover, the maximum and minimum velocity values move towards both sides of the cavities as the Strouhal number increases.

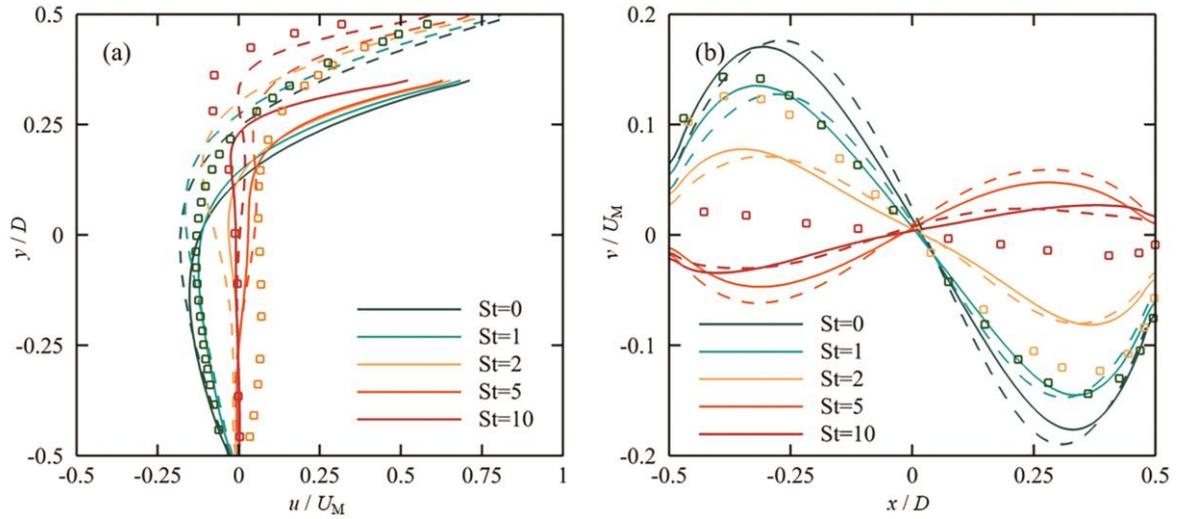

Figure 13. The profiles of normalized velocity for different cavity shapes (solid lines: P-Cavity, dashed lines: C-Cavity, and symbols: square cavity) and Strouhal numbers. (a) the profile of horizontal velocity component $u$ along the vertical line crossing the cavity center, and (b) the profile of vertical velocity component $v$ along the horizontal line crossing the cavity center. The results are shown at $t/t_s = 0$, and the Knudsen number is equal to $10^{-1}$. Data for square cavity is taken from Wu *et al.* [40].

## 5.2. Characteristics of the thermal field

### 5.2.1. Constant lid velocity

Figure 14 shows the predicted thermal field in cylindrical cavities with different shapes and Knudsen numbers, driven by a constant lid velocity. A non-uniform temperature distribution is observed in the cavities, with the left side experiencing a temperature decrease and the right side experiencing a temperature increase. When the Knudsen number is low, significant temperature variations occur only in the regions close to the corners of the moving lid, while the temperature remains relatively constant in the substantial part of the cavity. However, as the Knudsen number increases, temperature variations extend throughout the cavity. The results indicate that an increase in the Knudsen number leads to an increase in the temperature difference induced in the cavity. Moreover, the temperature difference in the C-Cavity is found to be larger than that in the P-Cavity.

As the moving lid drags the gas along, it generates a gas flow within the cavity that creates regions of high and low pressure in the cavity. In regions of lower pressure, the gas molecules experience a reduction in kinetic energy and subsequently slow down. This reduction in kinetic energy results in a decrease in the gas temperature, a phenomenon referred to as expansion cooling. The magnitude of the expansion cooling effect in a lid-driven cavity flow depends on the Knudsen number. At low Knudsen numbers ($Kn \ll 1$), the gas behaves like a continuum fluid, and the expansion cooling effect is negligible. However, as the Knudsen number increases, the gas becomes more rarefied, and the expansion cooling effect becomes more significant. Additionally, the C-Cavity experiences a more pronounced expansion cooling effect than the P-Cavity.

Interactions between gas molecules produce frictional forces, resulting in the dissipation of mechanical energy into thermal energy, known as viscous dissipation. This process plays a crucial role in altering the thermal field in rarefied lid-driven cavity flow with isothermal walls. The effect of viscous dissipation increases with increasing Knudsen number because the gas molecules are less likely to equilibrate with each other and transfer energy through collisions. Therefore, the mechanical energy input to the system from the moving lid is more likely to be dissipated into thermal energy through viscous dissipation, rather than being transferred to the gas molecules through collisions. Maximum viscous dissipation occurs where the largest magnitude of velocity gradient exists, which is predicted to be the top right corner, where the maximum shear stress occurs, making it the hottest region in the cavity. The magnitude of the velocity gradient generated in the C-Cavity is larger than that in the P-Cavity, resulting in a greater temperature rise in the former than the latter (see Figure 7(a)). A more notable temperature difference exists in the C-Cavity as compared to the P-Cavity, due to the greater expansion cooling effect and higher temperature increase caused by viscous dissipation in the former relative to the latter.

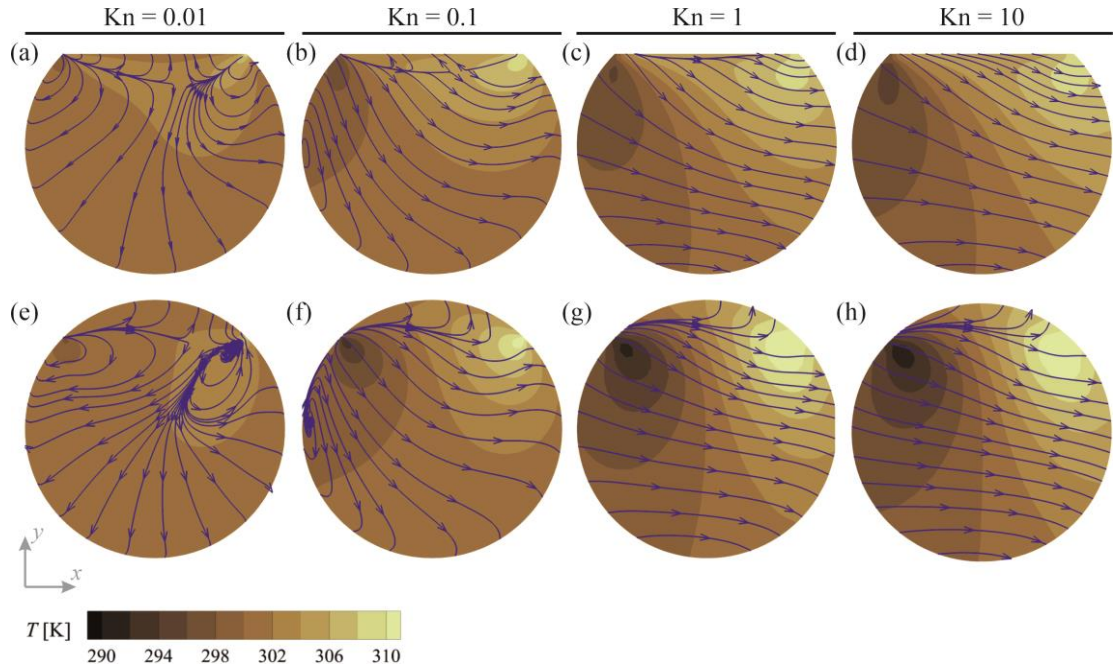

Figure 14. Contours of the temperature ($T$) overlayed by heat flux stream-traces for different cavity shapes and Knudsen numbers. The velocity magnitude of the cavity lid $U_{lid}$ is constant (*i.e.*, St = 0). Data corresponding to Knudsen numbers of 0.01 and 0.1 were acquired using the DUGKS model, while data for Knudsen numbers of 1 and 10 were obtained from the DSMC model.

The results presented in Figure 14 demonstrate that anti-Fourier heat transfer, also known as cold-to-hot heat transfer, occurs in the rarefied lid-driven cylindrical cavities even in cases where the Knudsen number is low (*i.e.*, Kn = $10^{-2}$). This phenomenon is attributed to the non-equilibrium nature of the rarefied gas, which cannot be described accurately by the Fourier heat conduction constitutive law based on the continuum theory [31, 48, 70]. The effects of rarefaction on the heat transfer behavior are accounted for in the gas kinetic theory [45]. In a rarefied fluid, the heat transfer from a hot region to a cold region is governed by the temperature gradient, while the heat transfer from a cold region to a hot region is governed by the gradient of shear stress, which is proportional to the second derivative of velocity with respect to space [70]. The results indicate that the contribution of the shear stress gradient to total heat transfer dominates that of the temperature gradient. Furthermore, an increase in the Knudsen number leads to an increase in the contribution of shear stress gradient to total heat transfer, resulting in predominantly cold-to-hot heat transfer, as depicted by the heat flux stream-traces in Figure 14. The presence of sharp corners in the P-Cavity causes a sudden change in the velocity field close to the corners, leading to an increase in the second derivative of velocity in this region. As a result, anti-Fourier heat transfer is more significant in the P-Cavity compared to the C-Cavity.

The predicted thermal field in cylindrical cavities with different shapes, employing various values for the tangential accommodation coefficients, is shown in Figure 15. The gas flow in the cavities is driven by a constant lid velocity and the Knudsen number is set to 1. The results reveal that changes in the tangential accommodation coefficient significantly affect the temperature distribution and the direction of heat flux in the cavity. A decrease in the tangential accommodation coefficient results in a reduction in the maximum temperature in the cavity and an increase in the minimum temperature, leading to a more uniform temperature distribution. Moreover, the heat flux stream-traces indicate that the heat flows predominantly from the moving lid towards the stationary bottom wall, with a decrease in the value of the tangential accommodation coefficient. This observation is attributed to the reduced effect of expansion cooling in thermal energy transfer in the cavity. A decrease in the tangential accommodation coefficient results in an increase in the number of molecules contributing to elastic energy exchange with the walls, leading to less wall shear stress and lower viscous heat dissipation [30].

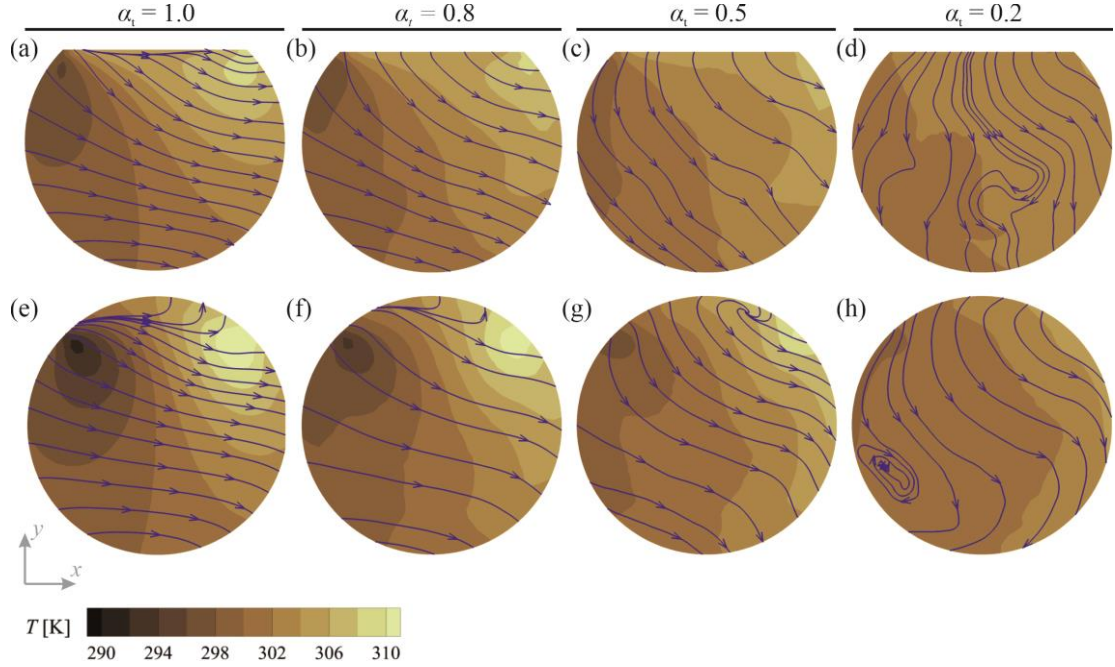

Figure 15. Contours of the temperature overlayed by heat flux stream-traces for different cavity shapes and tangential accommodation coefficients ($\alpha_t$). The velocity magnitude of the cavity lid $U_{lid}$ is constant, and the Knudsen number is equal to 1.

### 5.2.2. Oscillatory lid velocity

The contours of the temperature overlayed with heat flux stream-traces in cylindrical cavities are shown in Figure 16 for different Strouhal numbers at $t/t_s = 0$. For these cases, the Knudsen number is equal to $10^{-1}$. The results indicate that an increase in the lid oscillation frequency leads to an increase in the temperature difference induced in the cavity. In both cavity geometries, cold-to-hot heat transfer predominates when the Strouhal number equals 1. Interestingly, the heat flux stream-traces in the C-Cavity with St = 1 reveal the existence of a vortex heat flow structure near the cavity center. When the Strouhal number is increased to 2, it appears that the hot-to-cold heat transfer dominates in significant parts of the cavity, except in the vicinity of the corners of the lid. As the Strouhal number increases, the heat flow structure within the cavity becomes more intricate. Nevertheless, in substantial regions of the cavity, the hot-to-cold heat transfer continues to be dominant, except for the regions located proximate to the lid. The dominance of hot-to-cold heat transfer, particularly in the central region of the cavity, can be attributed to the increased viscous heat dissipation and, consequently, the elevated temperature gradient in the cavity. In regions proximate to the lid, the contribution of shear stress gradient to total heat transfer supersedes that of the temperature gradient, leading to the domination of cold-to-hot heat transfer near the cavity lid.

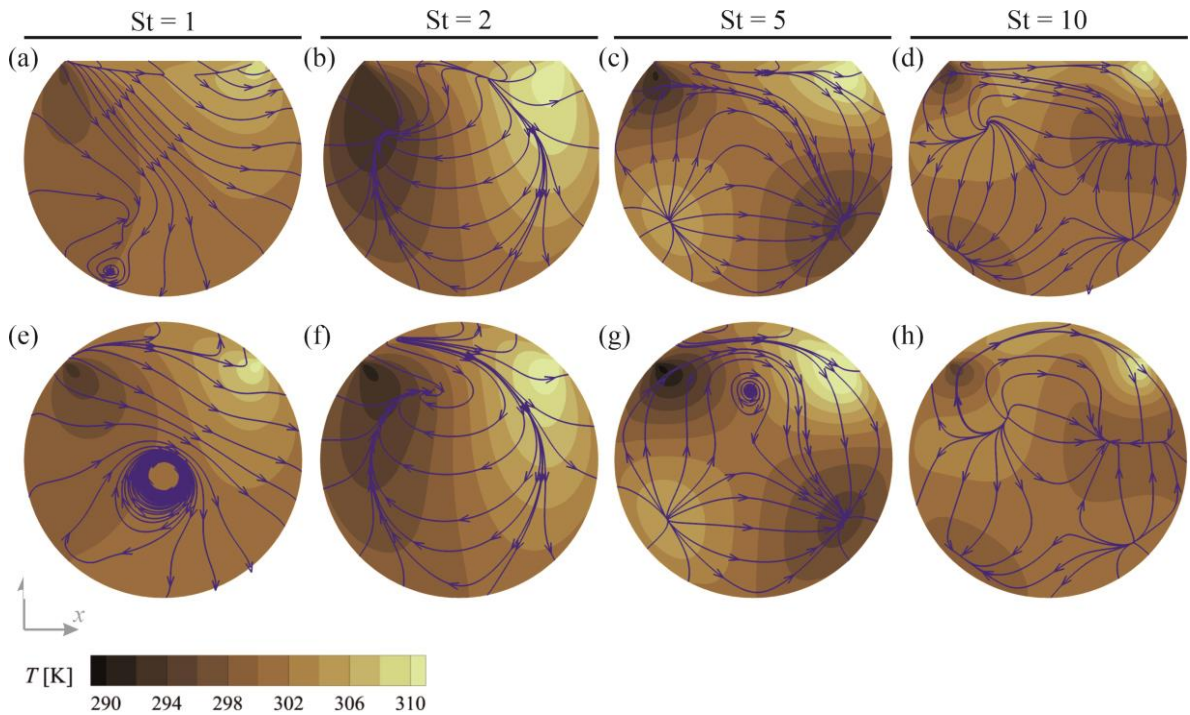

Figure 16. The contours of the temperature overlayed by heat flux stream-traces for different cavity shapes and Strouhal numbers. The results are shown at $t/t_s = 0$. The Knudsen number is equal to $10^{-1}$.

## 6. Conclusions and future directions

The rarefied gas flow and heat transfer in cylindrical lid-driven cavities with various cross-sections were studied numerically using both the direct simulation Monte Carlo (DSMC) and the discrete unified gas kinetic scheme (DUGKS) methods for a wide range of Knudsen numbers. The present study focused on understanding the effects of cavity geometry, degree of gas rarefaction, and boundary conditions on the thermal and fluid flow fields for constant and oscillatory lid velocities. The results were also compared with those obtained in square-shaped cavities.

The results obtained from the simulations suggest that the geometry of the cavity plays a significant role in influencing the characteristics of the thermal and gas flow fields. Specifically, the average gas velocity and induced temperature difference in cylindrical cavities with fully rounded edges (C-Cavity) are generally greater than that in cylindrical cavities with partially rounded edges (P-Cavity). Furthermore, it was observed that the expansion cooling and viscous dissipation are more pronounced in the C-Cavity compared to that in the P-Cavity. The results also revealed that anti-Fourier heat transfer, also known as cold-to-hot heat transfer, occurs in rarefied lid-driven cylindrical cavities with constant lid velocity, even when the Knudsen number is low (*i.e.*, Kn = $10^{-2}$). Moreover, anti-Fourier heat transfer is more significant in the P-Cavity compared to the C-Cavity.

The results indicate that changes in the tangential accommodation coefficient significantly affect the temperature distribution and the direction of heat flux in the cavity. A decrease in the tangential accommodation coefficient leads to a reduction in the maximum temperature in the cavity and an increase in the minimum temperature, resulting in a more uniform temperature distribution.

It was found that lid oscillations can reverse the heat flow direction from cold-to-hot to hot-to-cold, particularly in the central regions of the cavity. The outcomes of this study provide valuable insights into the behavior of rarefied gas flow and heat transfer in lid-driven cylindrical cavities with various cross-sections and can inform the design and optimization of relevant systems and devices.

To improve the computational efficiency of simulating rarefied gas flows in complex geometries, future research could investigate the development of hybrid numerical methods that combine the strengths of different approaches. Since the present work is limited to two-dimensional cylindrical cavities, further studies could explore three-dimensional rarefied gas flow instabilities in lid-driven cavities with complex geometries. Experimental investigations could also be conducted to verify the accuracy of the numerical simulations and shed light on the practical implications of the results for relevant engineering applications.